\documentclass[twocolumn]{aastex7}
\shorttitle{Understanding the magnetic field and plasma-$\beta$ along umbral fan loops}
\shortauthors{Rawat, Gupta et al.}
\begin{document}
\title{Understanding the magnetic field and plasma-$\beta$ along umbral fan loops traced using 3-min slow waves }
\author[orcid=0009-0005-9936-9928,sname='Rawat']{Ananya Rawat}
\affiliation{Udaipur Solar Observatory, Physical Research Laboratory, Dewali, Badi Road, Udaipur 313001, India}
\affiliation{Department of Physics, Indian Institute of Technology Gandhinagar, Palaj, Gandhinagar 382355, India}
\email[show]{ananyarawat@prl.res.in}  

\author[sname='Gupta']{Girjesh Gupta} 
\affiliation{Udaipur Solar Observatory, Physical Research Laboratory, Dewali, Badi Road, Udaipur 313001, India}
\email{girjesh@prl.res.in}

\author[sname=Van Doorsselaere]{Tom Van Doorsselaere}
\affiliation{Centre for mathematical Plasma Astrophysics, Department of Mathematics, KU Leuven, 3001 Leuven, Belgium}
\email{tom.vandoorsselaere@kuleuven.be}

\author[sname=Prasad]{S. Krishna Prasad}
\affiliation{Aryabhatta Research Institute of Observational Sciences (ARIES), Manora Peak, Nainital-263001, India}
\email{krishna.prasad@aries.res.in}

\author[sname=Erd\'elyi]{Robertus Erd\'elyi}
\affiliation{Solar Physics and Space Plasma Research Centre, School of Mathematical and Physical Sciences, University of Sheffield, Hicks Bldg, Hounsfield Road, Sheffield, S1 3RH, UK}
\affiliation{Department of Astronomy, E\"otv\"os Lor\'nd University, 1/A P\'azm\'any P\'eter s\'et\'any, H-1117 Budapest, Hungary}
\affiliation{Gyula Bay Zolt\'an Solar Observatory (GSO), Hungarian Solar Physics Foundation (HSPF), Pet\H{o}fi t\'er 3., Gyula, H-5700, Hungary}
\email{r.von.fay-siebenburgen@sheffield.ac.uk}

\begin{abstract}

The plasma-$\beta$ is an important fundamental physical quantity in solar plasma physics, which determines the dominating process in the solar atmosphere, i.e., magnetic or thermodynamic processes. Here, for the first time, we provide variations of magnetic field and plasma-$\beta$ along magnetically structured loops from the photosphere to the corona. We have selected several fan loops rooted in sunspot umbra observed simultaneously by the Interface Region Imaging Spectrograph and Solar Dynamics Observatory. The 3-min slow waves enabled us to trace and analyze several fan loops with cross-sectional areas in the lower atmosphere and locate their footpoints at the photosphere. We find the RMS magnetic field strengths in the range 1596-2269 G at the photospheric footpoints of the fan loops, which decrease rapidly to 158-236 G at the coronal footpoints. We estimated the plasma-$\beta$ at the photospheric and coronal footpoints in the range 0.2-0.5 and 0.0001-0.001, respectively. We found plasma-$\beta$$<$$1$ along the whole loop, whereas the plasma-$\beta$$\approx$$1$ layer is found to be at sub-photospheric heights. We compared our findings for isolated individual fan loops with a previously established model for active regions and found an almost similar pattern in variations with height, but with different plasma-$\beta$ values. Our results demonstrate the seismological potential of 3-min slow waves omnipresent in the umbral sunspot atmosphere to probe and map isolated loops and determine magnetic field and plasma-$\beta$ along these loops. The obtained parameters provide crucial ingredients for the theoretical modeling of the umbral atmosphere and wave dynamics along loops.   

\end{abstract}





\section{Introduction} 
\label{sec:intro} 

Sunspots on the surface of the Sun are intense collections of magnetic field lines or flux tubes, with near-vertical field strengths of $\approx 2000$ G at the photosphere \citep{2011LRSP....8....4B}. The signatures of these magnetic field lines extending upward into the corona are visible as loops in coronal images. Various structures and loops within the sunspots show evidence of propagating 3-min slow waves \citep{2015LRSP...12....6K}. Magnetic field strength decreases rapidly along the solar atmosphere and is difficult to measure at higher heights. However, oscillations in coronal loops have been used to estimate the magnetic field strength in the corona \citep[e.g.,][]{2001A&A...372L..53N,2008A&A...487L..17V,2008A&A...489L..49E,2009SSRv..149..229T,2016NatPh..12..179J}. Using MHD wave theory, \citet{2007ApJ...656..598W} determined magnetic field and plasma-$\beta$ in the range 21-51 G and 0.15-0.91, respectively or hot loops showing standing slow waves. Recently, \citet{2020Sci...369..694Y} measured the plane-of-sky component of the global coronal magnetic field to be 1-4 G at 1.05 to 1.35 $R_\odot$. 

\citet{2020ApJ...898L..34S} determined the average coronal magnetic field of 270$\pm$5 G using the spectroscopic data from Hinode. Similarly, \citet{2021ApJ...915L..24B} measured the coronal loop magnetic field strength and plasma-$\beta$ in the range 60-150 G and 0.0005-0.001, respectively. Coronal magnetic field strengths are also measured using magnetic field extrapolation techniques \citep[e.g.,][]{2009ApJ...696.1780D}. \citet{2020A&A...639A.114L} determined magnetic field and plasma-$\beta$ around 10 G and 0.02-0.1, respectively, along thin coronal loop emanating from non-sunspot region using the PFSS extrapolation technique.

Plasma-$\beta$ along the loops decides the potentiality of magnetic fields \citep{2001SoPh..203...71G} and characteristics of different waves present in the loops \citep{2015SSRv..190..103J}. \citet{2001SoPh..203...71G} developed a model for plasma-$\beta$ variation above active region from the photosphere to corona by combining various models and observational data. \citet{2004A&A...422..693M} utilized spectropolarimetric measurements and determined plasma-$\beta$ in the range of 0.5-1 inside the photospheric umbra. \citet{2013ApJ...779..168J} obtained the plasma-$\beta$=1 contour at the photospheric outer boundary of the sunspot penumbra using temperature and density from the sunspot model of \citet{1986ApJ...306..284M} and the magnetic field from HMI magnetogram. 
\citet{2017ApJ...850L..29B} used a 3D magnetohydrodynamic model of the solar corona over an active region to determine the plasma-$\beta$ from the photosphere to the corona. \citet{2017ApJ...837L..11C} solved analytical functions of the cut-off frequency and estimated the average plasma-$\beta$ value $\approx$0.83-0.86 within the umbra. 

Till now, all these results have been reported either in the global corona or only at certain loop segments. Variations of these parameters along the whole loops are still unexplored due to their non-traceability in the lower atmosphere and thus remain unclear.

Recently, \citet{2023MNRAS.525.4815R} demonstrated the unique technique using 3-min slow waves to trace loops from the corona to the photosphere via transition region and chromosphere in the umbral atmosphere along with their cross-sectional areas. Using these tracings, \citet{2023MNRAS.525.4815R,2024MNRAS.533.1166R,2024BSRSL..93..948R} studied the propagation and damping properties of slow waves along the loops for the first time from the photosphere to corona in detail. These findings can now also be utilized to estimate magnetic field strength and plasma-$\beta$ along loops from the photosphere to the corona, which was not possible before due to their non-traceability.

\begin{figure*}
    \centering
    \includegraphics[width=0.95\linewidth]{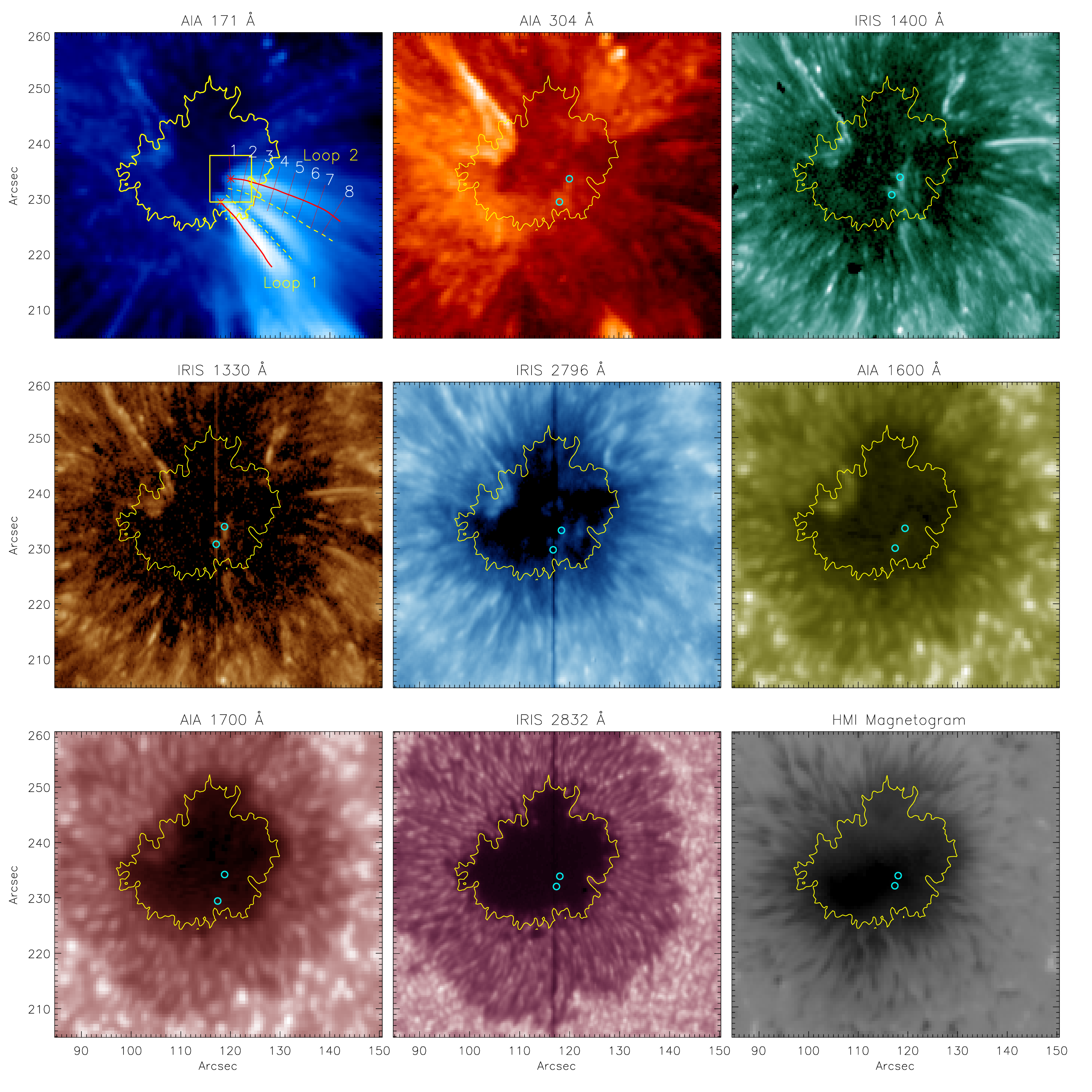}
    \caption{Images of sunspot and fan loops belonging to AR 12470 obtained from different AIA, IRIS, and HMI passbands as labeled. The red lines on the AIA 171 \AA\ image represent the manual tracing of coronal fan Loops 1 and 2, and asterisk symbols (*) represent their coronal footpoints. The yellow dashed lines represent the background regions for those loops. The sample slits across the coronal Loop 2 are marked with solid red lines and numbered. The box enclosing the coronal footpoint of Loop 2 indicates the region chosen to identify loop locations at the lower atmospheric heights. Small circles (o) over different panels represent the loop locations identified at that atmospheric height (see details in Section~\ref{sec:area}). Contours over different panels indicate the umbra–penumbra boundary as obtained from the IRIS 2832 \AA\ image.}
    \label{fig:img}
\end{figure*}

\subsection{Flux-tube theory\label{sec:tube}}

Loops visible in coronal images are manifestations of magnetic flux tubes. For a stable magnetic loop, the total pressure (thermal and magnetic) inside and outside should be balanced. For strong magnetic loops, the pressure balance equation is

\begin{equation}
\label{eq:balance}
  p_{int}(h)+\frac{B_{int}^2 (h)}{8\pi} = p_{ext}(h)+\frac{B_{ext}^2 (h)}{8\pi}
\end{equation} where $B_{int}$, $p_{int}$, $B_{ext}$ and $p_{ext}$ are the magnetic field and thermal pressure inside and outside the loop along loop length $h$, respectively. The thermal pressure decreases exponentially with height due to hydrostatic equilibrium. Therefore, the magnetic flux tube expands while keeping the total magnetic flux $\phi(h) = B(h)A(h)$ constant.

The ratio of the thermal plasma pressure ($p_{th}$) to the magnetic pressure ($p_m$) is known as plasma-$\beta$,
\begin{equation}
    \beta= \frac{p_{th}}{p_m}= \frac{8\pi N k_bT}{B^2}
\label{eq:beta}    
\end{equation}
where $k_b$ is the Boltzmann constant, $B$ is the magnetic field strength (G), $N=N_e+N_h$ is the total number density (cm$^{-3}$), and $T$ is the temperature (K), details in \citet{2001SoPh..203...71G}. Furthermore, semi-empirical models of solar atmosphere developed by \citet{1981ApJS...45..635V,1986ApJ...306..284M,1999ApJ...518..480F} can provide $N_e$, $N_h$, and $T$ variations with height.

In this letter, for the first time, we provide estimates on the magnetic field and plasma-$\beta$ variations along the various fan loops traced from the photosphere to the corona using 3-min waves and along the corona.

\section{Observations}
\label{sec:obs}

To determine the variations of magnetic field and plasma-$\beta$ along the fan loops rooted within the sunspot umbra, we need to trace the fan loops in the lower atmosphere (photosphere to low corona) where loops are not visible. These loops can be traced in the lower atmosphere using 3-min slow waves \citep[details of technique is provided in][]{2023MNRAS.525.4815R}. For this purpose, we have identified an appropriate dataset observed by the Atmospheric Imaging Assembly \citep[AIA;][]{2012SoPh..275...17L}, the Helioseismic and Magnetic Imager \citep[HMI;][]{2012SoPh..275..207S} both onboard the Solar Dynamics Observatory \citep[SDO;][]{2012SoPh..275....3P}, and the Interface Region Imaging Spectrograph \citep[IRIS;][] {2014SoPh..289.2733D}. To obtain good coverage over the lower atmosphere, we have identified an active region that was observed by all four IRIS slit jaw images (SJIs). This makes the tracing of the loops in the lower atmosphere more robust. The sunspot studied here belongs to the Active Region (AR) 12470, observed on 2015 December 19. We obtained 55 min of simultaneous data starting from 13:26:22 UT, as shown in Figure~\ref{fig:img}. We also utilize data from the sunspot belonging to AR 12553 observed on 2016 June 16, in which several fan loops were already traced and analyzed by \citet{2023MNRAS.525.4815R,2024MNRAS.533.1166R,2024BSRSL..93..948R}. Here, we selected a few quiescent loops for our analysis as presented in Appendix~\ref{Appendix:fanloop}.

Images obtained from the IRIS-SJI passbands have an exposure time of 2 s with an effective cadence of 12.75 s, 0.332 arcsec pixel$^{-1}$ resolution, and 169''$\times$182'' field of view. The analyzed sunspot is slightly off the disc center (heliocentric coordinates X $\approx$ 115'', Y $\approx$ 235''), the angle between the local vertical and the line-of-sight is $\approx 25^o$, which leads to $\mu$ = cos $\theta \approx$ 0.90. Therefore, we have ignored projection effects on intensity oscillations and other parameters. 

We co-aligned the AIA, IRIS, and HMI datasets using the IRIS-SJI 2796 \AA\ and AIA 1700 \AA\ image pair and the IRIS-SJI 2832 \AA\ and HMI continuum image pair by utilizing the cross-correlation method. AIA, IRIS, and HMI images are derotated with respect to starting time of IRIS using SSW routines. The identified dataset provides a unique opportunity to study atmospheric seismology through wave propagation along the whole solar atmosphere.

\citet{2006SoPh..239...69N} estimated the formation height of the HMI from Fe I 6173 \AA\ line. They derived the HMI continuum formation height from line continuum to be $\approx$21 km, and HMI Doppler and magnetogram formation height from the line core to be $\approx$ 269 km above the optical depth of unity ($\tau_{5000}$=$1$). These formation heights correspond to temperatures around 4200 K and 3700 K, respectively \citep{1999ApJ...518..480F}. AIA 171, 304, 1600, and 1700 \AA\ passbands correspond to a coronal temperature$\approx$ 0.8 MK, transition region temperature $\approx$0.05 MK, chromospheric temperature $\approx$5700 K and lower chromospheric temperature$\approx$ 4500 K, respectively. 

IRIS-SJI-1330 and 1400 \AA\ passbands correspond to transition region temperatures$\approx$ 20,000 K and 63,000 K, derived from C II and Si IV spectral lines, respectively. IRIS-SJI-2796 \AA\ corresponds to a chromospheric temperature$\approx$ 10,000 K derived from the Mg II, and IRIS-SJI-2832 \AA\ corresponds to a photospheric temperature$\approx$ 4200 K derived from the photospheric continuum \citep{2014SoPh..289.2733D}.

\section{Data Analysis and Results}
\label{sec:analysis}

Figure~\ref{fig:img} shows the analyzed fan loop system rooted in sunspot umbra in AR 12470 in AIA 171 \AA\ image. The overplotted contour represents the umbral boundary identified from the IRIS-SJI-2832 \AA\ passband. Fan loops are manually traced on the AIA 171 \AA\ image, and asterisk (*) symbols represent the loop footpoints in the corona. We analyze several clean loops from AR 12470 and 12553. However, here we present results from Loop 2 from AR 12470 as a representative example, due to its 40\% longer length and lower background signal, which allows estimation of loop cross-sections up to longer distances as compared to Loop 1. Section~\ref{sec:beta} summarizes the results from all the selected loops emanating from the sunspot umbra in AR 12470 and 12553.

\begin{figure*}
    \centering
    \includegraphics[width=0.95\linewidth]{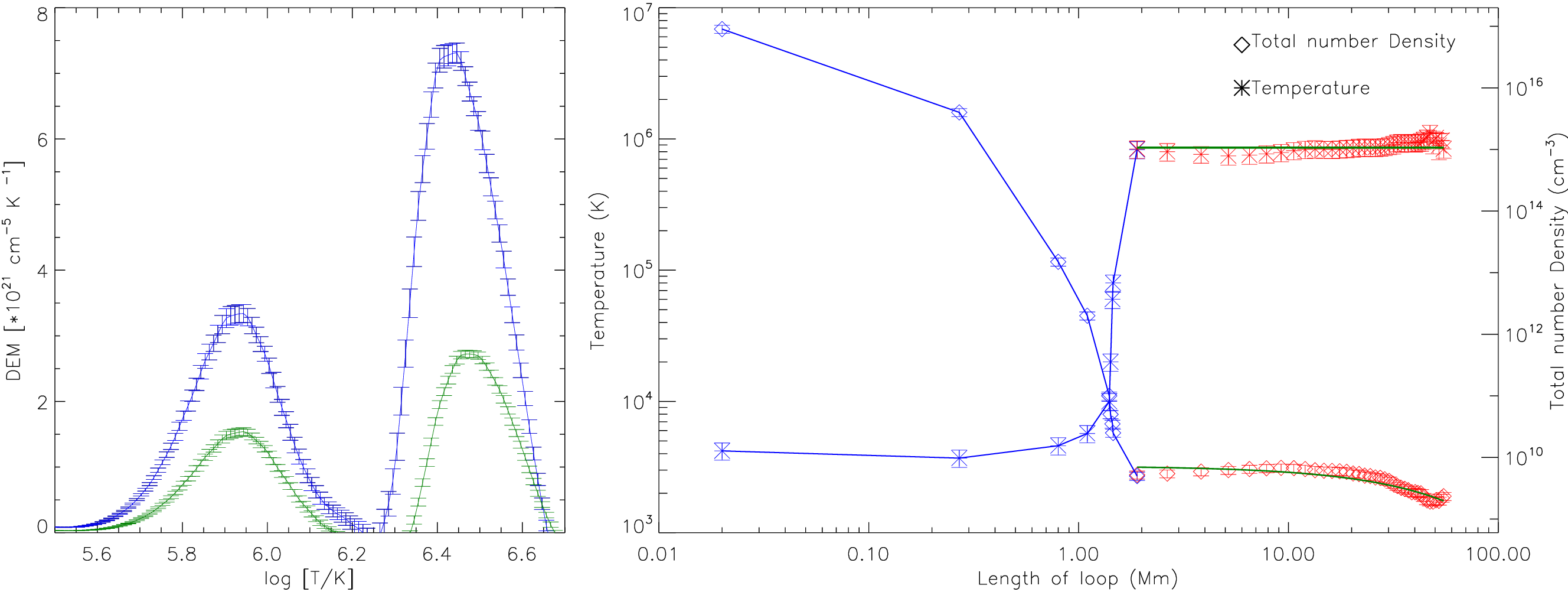}
    \caption{Left: DEM profiles at the coronal footpoint of the Loop 2 shown by the asterisk in Figure~\ref{fig:img}. The green and blue curves are obtained with and without background-subtracted intensities, respectively. Right: Temperature (left $y$-axis) and total number density (right $y$-axis) as a function of loop length shown by asterisks (*) and diamonds ($\diamond$), respectively. Temperature and density from the photosphere to the corona are obtained from the sunspot model of \citet{1999ApJ...518..480F} and shown in blue lines, whereas those in the corona are obtained using DEM analysis and shown in red symbols. The green lines overplotted over coronal temperature and density represent average loop temperature and a fitted exponentially decaying function, respectively.}
    \label{fig:temp_den}
\end{figure*}

\subsection{Temperature and density along the loop}
\label{sec:temp}

Since the temperature sensitivity of each IRIS and SDO passband is well known (see Section~\ref{sec:obs}), we deduce the formation height and thus total number density ($N=N_e + N_h$; plasma in the lower atmosphere are partially ionized) corresponding to each IRIS and SDO passband by utilizing the sunspot model of \citet{1999ApJ...518..480F}, and plotted in Figure~\ref{fig:temp_den}. In this model, the $0$ km height represents the $\tau=1$ layer at 5000 \AA\ in the sunspot umbra.

To determine the $T$ and $N_e$ along the loop in the corona, we carried out differential emission measure (DEM) analysis using the tool provided by \citet{2012A&A...539A.146H} on the background-subtracted intensities. Sample DEM profiles at coronal footpoint are plotted in the left panel of Figure~\ref{fig:temp_den}. The first DEM peak around $0.85$ MK represents the emission from the fan loops \citep[e.g.,][]{2017ApJ...835..244G,2018ApJ...868..149K}. 
We obtained the $N_e$ along the coronal loop using EM values at peak temperature using $N_e$=$\sqrt{(EM/w)}$, where $w$ is the coronal loop diameter as deduced in Section~\ref{sec:area}. We assume the plasma filling factor across the loop to be 1. The obtained $T$ and $N$ ($N$=$N_e + N_h$ $\approx 2N_e$; coronal plasma is completely ionized) along the coronal loop are plotted in Figure~\ref{fig:temp_den}. The temperature along the loop slightly increases, although this increase is well within the error bars. The average temperature along the loop is 0.86 MK. Since the density obtained from the DEM method ($5\times10^9$ cm$^{-3}$) at the coronal footpoint is slightly different from the density obtained from the sunspot model \citep[$10^{10}$ cm$^{-3}$,][]{1999ApJ...518..480F}, we scaled the density of the sunspot model by dividing it with 2 in Figure~\ref{fig:temp_den} to match the density at the coronal footpoint. Furthermore, we have assumed a 15\% error in density and temperature in all our calculations \citep[e.g.,][]{2019A&A...627A..62G}. The blue and red colors represent variations in the lower atmosphere and corona, respectively, and the same color code will be used throughout. 

\subsection{Loop inclination}

We fitted the total number density variation along the loop in the corona with an exponential decay function shown with the green line in the right panel of Figure~\ref{fig:temp_den}. The fit provides observed density scale height $\lambda_{d\_obs} \approx$ 12.9$\pm$0.73 Mm along the coronal loop. This is a plane-of-sky projected scale height due to the inclination of loop. All the fits are performed using the MPFITFUN routine \citep{2009ASPC..411..251M}. We compare this $\lambda_{d\_obs}$ with the expected density scale height ($\lambda_{d\_exp}$) from hydrostatic equilibrium in corona, i.e., $\lambda (T)= \frac{k_b T}{ \mu m_p g} \approx 46 [\frac{T}{1 MK}]$ Mm, where $\mu$ is mean molecular weight, $m_p$ is the mass of the proton, and $g$ is the acceleration due to gravity at the solar surface \citep[see e.g.,][]{1999ApJ...515..842A,2015ApJ...800..140G}. Since the average temperature of the loop is 0.86$\pm$0.13 MK, $\lambda_{d\_exp}\approx39.56$$\pm$5.98 Mm. This difference in both scale heights provides the inclination angle ($\phi$) of the loop, \begin{equation}
    \cos(\phi)= \frac{\lambda_{d\_obs}}{\lambda_{d\_exp}}
\end{equation} \citep[e.g.,][]{1999ApJ...515..842A}. The loop is $\phi = 71^{\circ}$ inclined with respect to the plane-of-sky. Therefore, we have corrected the coronal loop length by multiplying it with a factor of 3.07.

\subsection{Loop locations and cross-section areas at different heights}
\label{sec:area}

\begin{figure*}
    \centering
    \includegraphics[width=0.99\linewidth]{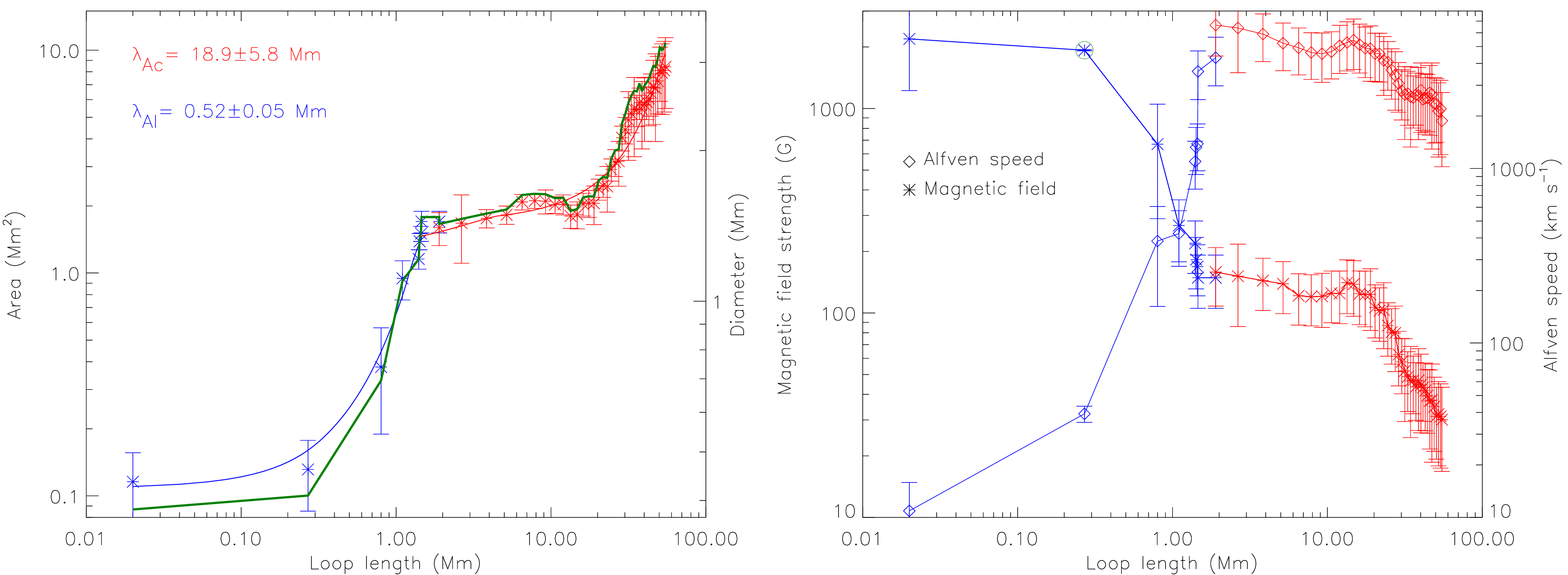}
    \caption{Left: Cross-sectional area of the loop along its length. The area of the loop in the lower solar atmosphere is fitted with an exponentially rising function, shown in a blue solid line. The area along the coronal loop is also fitted with an exponentially rising function with a constant background, shown in the red solid line. Obtained area scale heights ($\lambda_A$) are printed in their respective color code at the top left corner. The green line represents the equivalent diameter ($w$) of the loop along its length (right axis). Right: Variation of magnetic field strength (left axis) and Alfv\'en wave speed (right axis) along the loop. The green circle ($O$) represents the RMS magnetic field strength obtained from the HMI magnetogram.}
    \label{fig:area-speed}
\end{figure*}

To determine the diameter and cross-sectional area along the coronal fan loops in the AIA 171 \AA\ image, we obtained the intensity profile along several slits across the loop. We fitted the intensity profiles with the Gaussian function and a linear background \citep[e.g.,][]{2019A&A...627A..62G}. We then extracted the full width at half maximum (FWHM) of the fitted Gaussian as the diameter of the loop. Obtained diameter at coronal footpoint is $\approx$ 3.28$\pm$0.28 pixel. Sample intensity profiles and fittings at a few slit locations, marked in Figure~\ref{fig:img}, are presented in Appendix~\ref{Appendix:fwhm}.

For lower heights, where the loops are not visible, recently \citet{2023MNRAS.525.4815R} devised a technique to trace back the footpoint and cross-section area of loops on the photosphere through the transition region and chromosphere. As per the method, we choose a box of size 9"$\times$9"  by positioning the coronal footpoint of the loop in the center (a sample box chosen for loop 2) is shown in Figure~\ref{fig:img}. We performed a correlation analysis using 3-min filtered light curves and obtained correlation images at different heights; see details in Appendix~\ref{Appendix:area}. 

A contour level of about 92\% of the maximum correlation coefficient value on the correlation image for the loop footpoint in the corona matches fairly well to the loop cross-section obtained from the FWHM. At various other heights in the lower atmosphere, closed contours are obtained within $94\pm2$ \%  of the maximum correlation values. The pixels with the maximum correlation coefficients in the correlation images depict the location of the loop at that atmospheric height. These locations are marked as small circles in Figure~\ref{fig:img}.

Further, we take the sum of the area of all the pixels falling within the contour of $\approx$$94\pm2\%$ of the maximum correlation value of the correlation image between two atmospheric heights. This area can be considered as the representative size of the loop, i.e., the loop cross-section. The area within this correlation threshold decreases as we move into the lower atmosphere. This indicates loop expansion with increasing height in the lower atmosphere and suggests the geometric spreading of 3-min slow waves along the expanding loop. Furthermore, we determine the area scale height in the lower atmosphere by fitting the obtained cross-sectional area with an exponentially rising function
\begin{equation}
    A(h)= A_o e^{\frac{h}{\lambda_A} }.
\label{eq:area}
\end{equation}

Here, $A(h)$ is the cross-sectional area at loop length $h$, $\lambda_A$ is the area scale height, and $A_o$ is an appropriate constant. The obtained area scale height for the lower atmosphere is $\lambda_{Al}$$\approx 518\pm 46$ km. 

To obtain the cross-sectional area along the loop in the corona, we calculated $\pi(\frac{FWHM}{2})^2$. The cross-sectional area along the loop is increasing, as shown in Figure~\ref{fig:area-speed}. This expansion can also be visualized in the AIA 171 \AA\ image in Figure~\ref{fig:img}. The obtained area of the loop footpoint in the corona is 3.04$\pm$0.26 arcsec$^2$ from the FWHM method and 3.24$\pm$0.18 arcsec$^2$ from the correlation analysis. Both values are well within the error bar. We fitted the area expansion along the coronal fan loop using an exponentially rising function with a constant background and obtained scale height $\lambda_{Ac}$ $\approx18.89\pm5.77$ Mm.

The cross-sectional area shown in Figure~\ref{fig:area-speed} is as per the expectations from the theory of flux tube expansion with height, where expansion in the lower atmosphere is faster (smaller $\lambda_A$) than expansion along the corona (larger $\lambda_A$). This is because the pressure scale height in the lower atmosphere is much smaller than that in the corona due to the steep change in atmospheric temperature. We also provide the diameter ($w$) of the loop in Figure~\ref{fig:area-speed}, derived from the loop cross-sectional area upon assuming equivalent cylindrical loop geometry ($w=2\sqrt{\frac{A(h)}{\pi}}$). 

\subsection{Magnetic field strength and plasma-$\beta$ along the fan loop}
\label{sec:beta}

Loops are a manifestation of magnetic flux tubes in which the total magnetic flux remains constant along its length. We have already obtained the cross-sectional area of the loop at the photosphere ($A_p \approx$ 0.22 arcsec$^2$) and its variation along the loop length ($A(h)$) as shown in Figure~\ref{fig:area-speed}. From the loop tracing, we have identified the loop footpoint at the photosphere and obtained the magnetic field using the HMI magnetogram data (at atmospheric height 269 km). We determined the root mean square (RMS) magnetic field strength ($B_p$) with one sigma error at the photosphere as $1928\pm13$ G during the 55-min time sequence. The magnetic field strength along the loop length is calculated using
\begin{equation}
\label{eq:mag}
    B(h)= \frac{B_pA_p}{A(h)}.
\end{equation}
The error in $B(h)$ is calculated from errors in the cross-sectional area and photospheric magnetic field strength. The obtained magnetic field variation with the error bars along the loop length is plotted in Figure~\ref{fig:area-speed}. We obtained the magnetic field strength at the coronal footpoint $\approx 158\pm$50 G. Furthermore, we also determined the Alfv\'en speed variation along the loop length using,
\begin{equation}
    V_A(h)= \frac{B(h)}{\sqrt{4\pi\rho(h)}}
    \label{eq:speed}
\end{equation}
where $\rho(h)=N_e(h)m_e+N_h(h)m_h$ is the total mass density and plotted in Figure~\ref{fig:area-speed}. Alfv\'en speed increases in the lower atmosphere, peaks at coronal footpoint, and then decreases further in the corona. Such speed variation are also noted before \citep[e.g.,][]{1999ApJ...515..842A,2012ApJ...754...92C}. 

\begin{figure*}
    \centering
    \includegraphics[width=0.80\linewidth]{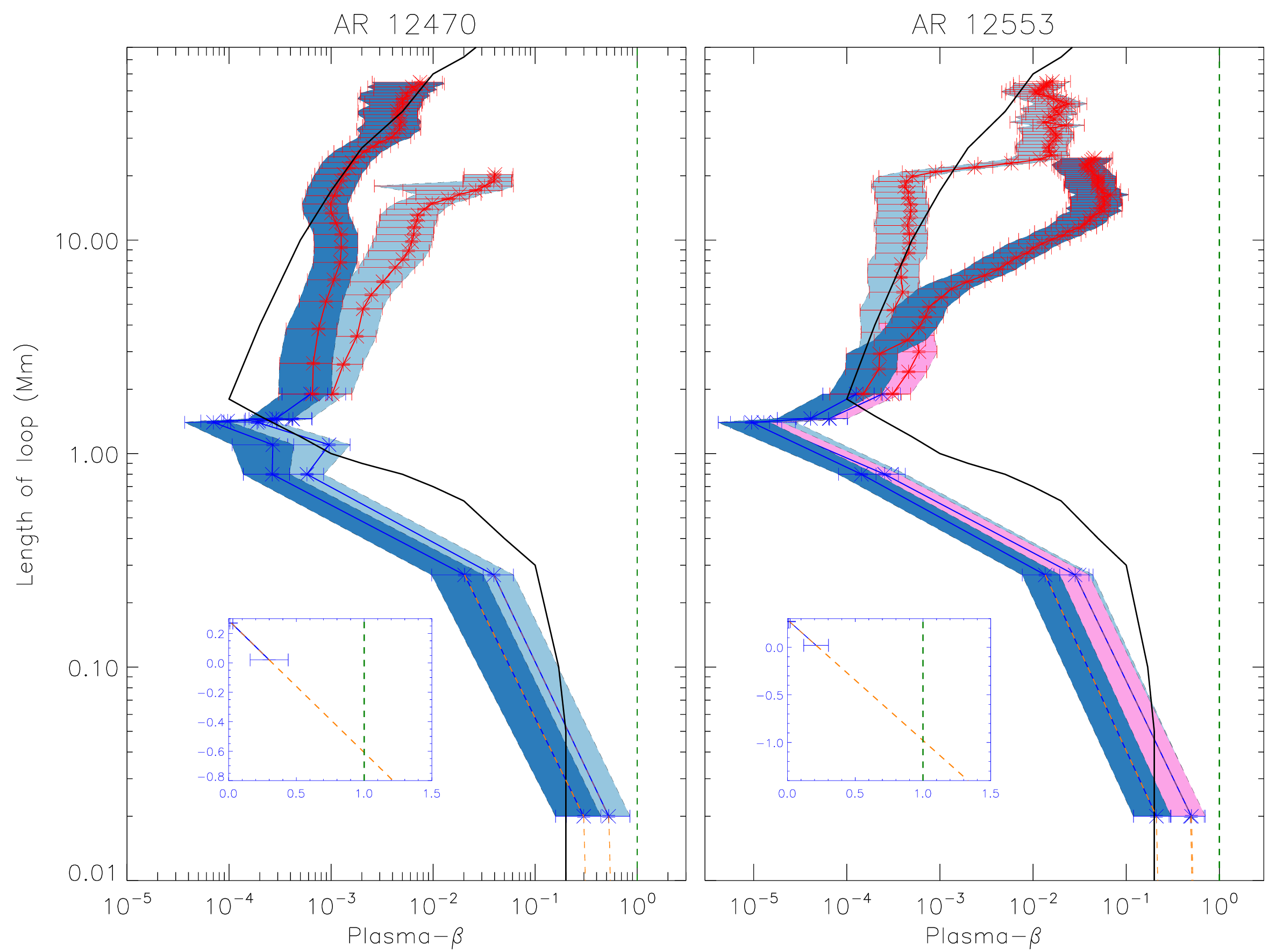}
    \caption{Left: Variation of plasma-$\beta$ along the fan loops belonging to AR12470 where light and dark blue shaded regions are for loops 1 and 2, respectively. Right: Variation of plasma-$\beta$ along the fan loops belonging to AR12553 where the light blue, dark blue, and pink shaded regions are for loops 3, 5, and 6, respectively. The black solid lines represent data points extracted from \citet{2001SoPh..203...71G} for the umbral region. The vertical green dashed lines indicate the plasma-$\beta = 1$ layer. The orange dashed lines are the linear extrapolation of data points to determine the plasma-$\beta = 1$ layer. The small boxes in the lower left corners show the height at which loops 2 and 5 cross the plasma-$\beta$$\approx1$ layer as representative examples.}
    \label{fig:beta}
\end{figure*}

\begin{table*}

    \caption{Various parameters derived along fan loops identified in AR 12470 and 12553. $\lambda_{Al}$ and $\lambda_{Ac}$ are the area scale height of loops from the photosphere to the transition region and in the corona, respectively. $B_p$ and $B_c$ are magnetic fields at photospheric and coronal footpoints of fan loops, respectively. $<T>$ and $N_{cf}$ are the average coronal loop temperature and total number density at the coronal footpoint, respectively. $V_{Ac}$ is the Alfv\'en speed at coronal footpoints. $\beta_p$ and $\beta_c$ are the plasma-$\beta$ values at photospheric and coronal footpoints, respectively. The plasma-$\beta$$\approx1$ heights along fan loops are measured with respect to the optical depth of unity ($\tau_{5000}=1$).}
     \label{tab:stat}
     \centering 
      \resizebox{\textwidth}{!}{
      {\begin{tabular}{|c|c|c|c|c|c|c|c|c|c|c|c|c|c|c|c|c|c|}
   \hline
 Loop no. & $\lambda_{Al}$ & $\lambda_{Ac}$ & $B_p$ & $B_c$ &$<T>$ & $N_{cf}$ & $V_{Ac}$  & $\beta_p$  & $\beta_c$  & $\beta$$\approx1$ \\

 &  (km) &  (Mm)& (G) & (G)  & (MK) &cm$^{-3}$ & (km s$^{-1}$) &  &  &  height (km)\\ \hline
AR 12470     \\ \hline

1  & 521$\pm$44 & 5.85$\pm$1.85 & 1828$\pm$12 & 164$\pm$59 & 0.85& 7.8$\times$10$^9$& 5173 & 0.52 & 0.001  & -227 \\ \hline
 2 & 518$\pm$46 & 18.89$\pm$5.77& 1928$\pm$13 & 158$\pm$50 &0.86 &5.0$\times$10$^9$ & 6637 & 0.30 & 0.0006 & -617  \\ \hline
AR 12553  \\ \hline
3  & 884$\pm$152 &4.08$\pm$0.55 & 1596$\pm$15  & 236$\pm$86&0.86 &5.2$\times$10$^9$ & 10004 & 0.49 & 0.0002 & -264 \\ \hline
5  & 769$\pm$114 & 4.22$\pm$1.1 & 1907$\pm$120 & 224$\pm$82&0.82 &3.0$\times$10$^9$ & 12209 & 0.21 & 0.0001  & -980 \\ \hline
6  & 900$\pm$168 & --           & 2269$\pm$06  & 227$\pm$83&0.89&5.1$\times$10$^9$ & 9146 & 0.50 & 0.0003  & -244 \\ \hline
    \end{tabular}}}
\end{table*}

The obtained density, temperature, and magnetic field strength are further utilized to estimate the plasma-$\beta$ along the loop using Equation~\ref{eq:beta}. The results are presented in Figure~\ref{fig:beta}. The dark blue shaded region in the left panel represents the variation of plasma-$\beta$ along Loop 2. In the bottom left panel, we fitted a straight line between the temperature minimum region and photospheric plasma-$\beta$ values. We extended it further to find the plasma-$\beta$$=1$ layer at $\approx-617$ km height, as shown in the orange dashed line. The negative height signifies that the plasma-$\beta$$=1$ layer is below the photosphere ($\tau_{5000}$$=1$). The plasma-$\beta$ at the photospheric and coronal footpoints are $0.3\pm0.14$ and $0.0006\pm0.0003$, respectively. Additionally, as we move along the loop in the corona, the loop merges with the background corona, making it difficult to measure any loop parameters beyond 60-70 Mm. Moreover, density estimation from DEM analysis may change depending upon background subtraction location; accordingly, plasma-$\beta$ estimates in the corona will also change. However, such changes will be well within the estimated error bars of about 50\%. 

We performed a similar analysis on the other fan loops identified in AR12470 (see Figure~\ref{fig:img}) and also in AR12553 (see Appendix~\ref{Appendix:fanloop}), where loops were already traced by \citet{2023MNRAS.525.4815R}. All the loop footpoints at the photosphere are directed toward the umbral center, and their cross-sectional areas are smaller than their coronal counterparts, as found for Loop 2. The results suggest more or less similar statistics for all the loops in the lower atmosphere and corona with similar plasma-$\beta$ variations. The summary of all the analyzed loops is provided in Table~\ref{tab:stat}.

\section{Summary}
\label{sec:summary}

In this work, we traced several umbral fan loops in the corona and from the corona to the photosphere, where loops are not visible. We obtained their cross-sectional area at different heights in the solar atmosphere and located their photospheric footpoints. The area of Loop 2 decreased from 3.24$\pm$0.36 arcsec$^2$ (1.70$\pm0.19$ Mm$^2$) in the corona to 0.22$\pm0.11$ arcsec$^2$ (0.11$\pm0.05$ Mm$^2$) at the photosphere. The scale height of area expansion for loop 2 from the photosphere to the corona is $\approx0.52$ Mm, whereas that in the corona is $\approx19$ Mm. Such expansions are expected in the solar atmosphere (see Section~\ref{sec:tube}). Similar loop expansion from photosphere to corona is also noted by \citet[][and references therein]{2025ApJ...980L..20B}. 

We estimated the magnetic field strength of $158\pm50$ G at the coronal loop footpoint that decreases slowly along the coronal loop. Our estimates match very well with the magnetic field values reported by \citet{2021ApJ...915L..24B} using the recently developed spectroscopic diagnostic technique and also with the measurement of \citet{2022MNRAS.512.3149G} of about 143 G along the transient loop using the magnetic field extrapolation method. These coronal magnetic field measurements along the loops are much larger than the usually quoted average global coronal magnetic field strength of 1-4 G \citep{2004ApJ...613L.177L,2020Sci...369..694Y}, and also the average coronal loop magnetic field of 4-30 G obtained from kink oscillations \citep{2001A&A...372L..53N,2011ApJ...736..102A}. Such large estimates of the magnetic fields along the loops will lead to larger Alfv\'en wave energy flux than those quoted in literature \citep[e.g.,][]{2009A&A...501L..15B,2011Natur.475..477M,2019A&A...627A..62G} as needed in Alfv\'en wave heating models \citep[e.g.][]{2018NatPh..14..480G,2024A&A...688A..80K}. 

Plasma-$\beta$ variations reported by \citet{2001SoPh..203...71G} for the umbral regions are similar to those obtained for fan loops with slightly different values (see Figure~\ref{fig:beta}). Interestingly, plasma-$\beta$ values at chromospheric heights are consistently smaller than those provided in \citet{2001SoPh..203...71G}. At longer loop lengths, fan loops merge with the active region background, and our plasma-$\beta$ values also merge with \citet{2001SoPh..203...71G}. The differences in the values could be due to the fact that we are tracing single isolated loops, whereas the values extracted from \citet{2001SoPh..203...71G} are for integrated sunspot umbra. The overall similarity with the variation pattern of \citet{2001SoPh..203...71G} indicates that our loop tracing method works well in the lower atmosphere. The value of plasma-$\beta$$<$1 along the whole loop length from the photosphere to the corona indicates that the umbral loops are potential in nature \citep[e.g.,][]{2011LRSP....8....4B}. We also noted that the plasma-$\beta$$=$$1$ layer exists at sub-photospheric heights \citep[e.g.,][]{1994ApJ...437..505C}. 
This layer is important for helioseismological studies \citep[e.g.,][]{2015LRSP...12....6K} and demands a detailed investigation. 

In summary, we obtained variations of cross-sectional area expansion, magnetic field strength, and plasma-$\beta$ along the individual loops from the photosphere to the corona. Estimated parameters will provide useful ingredients for the MHD modeling of loops emanating from the sunspot regions and wave dynamics in the density-stratified and expanding waveguides \citep[e.g.][]{2009SSRv..149..229T,2012ApJ...748..110L}. 
It will also be important to carry out detailed statistical investigations of the estimates on magnetic field and plasma-$\beta$ along loops emanating from different regions on the Sun and their comparison with different techniques.  

\begin{acknowledgments}
We thank the referee for the helpful suggestions, which improved the quality of presentation. The research work at the Physical Research Laboratory (PRL) is funded by the Department of Space, Government of India. AR thanks PRL for her PhD research fellowship. TVD received financial support from the Flemish Government under the long-term structural Methusalem funding program, project SOUL: Stellar evolution in full glory, grant METH/24/012 at KU Leuven. The research that led to these results was subsidised by the Belgian Federal Science Policy Office through the contract B2/223/P1/CLOSE-UP. It is also part of the DynaSun project and has thus received funding under the Horizon Europe programme of the European Union under grant agreement (no. 101131534). Views and opinions expressed are however those of the author(s) only and do not necessarily reflect those of the European Union and therefore the European Union cannot be held responsible for them. SKP is grateful to SERB/ANRF for a startup research grant (No. SRG/2023/002623). R.E. acknowledges the NKFIH (OTKA, grant No. K142987) Hungary for enabling this research.  R.E. is also grateful to Science and Technology Facilities Council (STFC, grant No. ST/M000826/1) UK, PIFI (China, grant No. 2024PVA0043) and the NKFIH Excellence Grant TKP2021-NKTA-64 (Hungary). This work was also supported by the International Space Science Institute project (ISSI-BJ ID 24-604) on "Small-scale eruptions in the Sun". AIA and HMI data are courtesy of SDO (NASA). IRIS is a NASA small explorer mission developed and operated by LMSAL with mission operations executed at NASA Ames Research Center and major contributions to downlink communications funded by the Norwegian Space Center (NSC, Norway) through an ESA PRODEX contract.    
\end{acknowledgments}


\appendix

\section{Fan loop system belonging to AR 12553}
\label{Appendix:fanloop}
\begin{figure}
    \centering
    \includegraphics[width=0.60\linewidth]{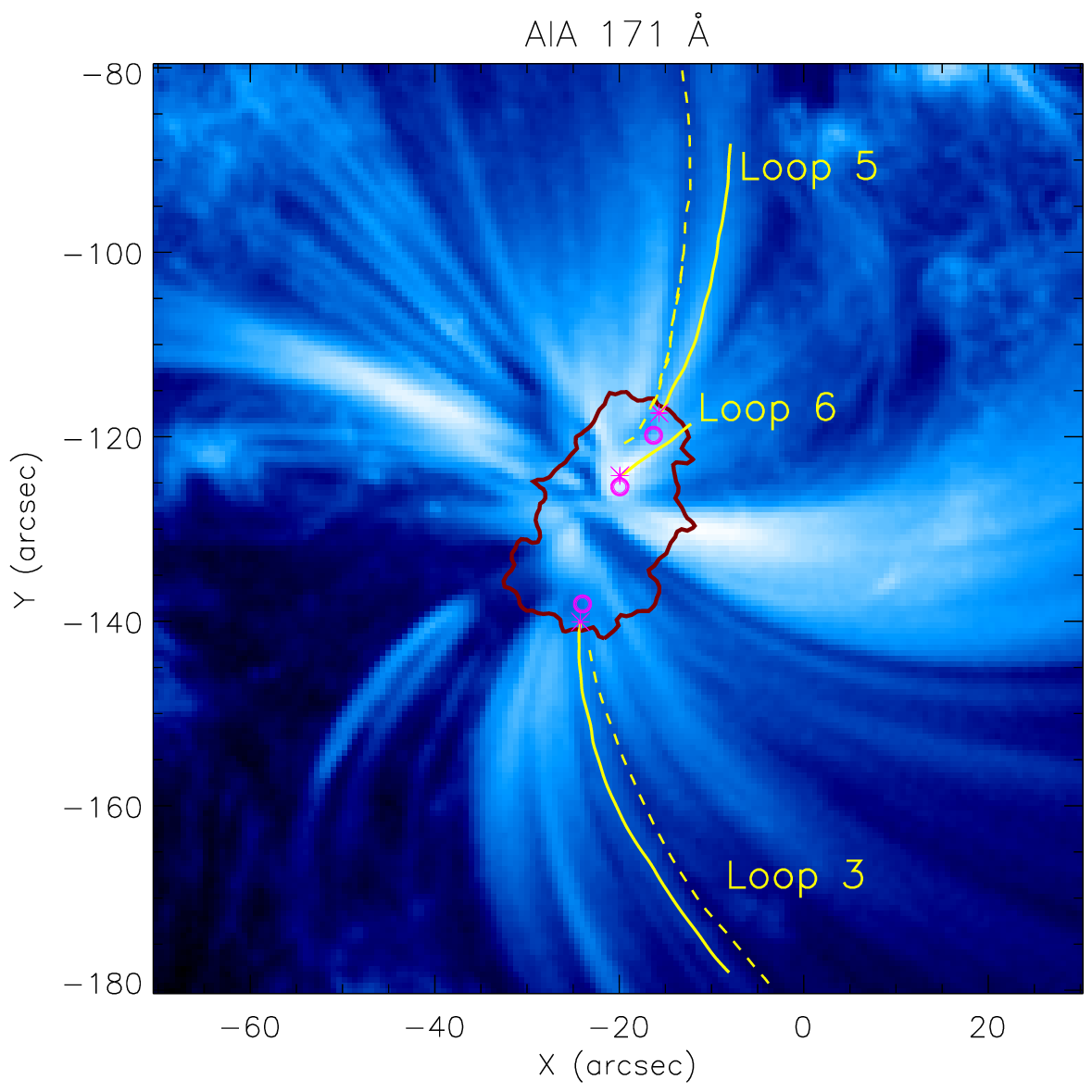}
    \caption{Image of the fan loop system belonging to AR 12553 obtained from AIA 171 \AA\ passband studied in \citet{2023MNRAS.525.4815R}. The yellow solid lines represent the manual tracing of coronal Loops 3, 5, and 6 as labeled, and asterisk symbols (*) represent their coronal footpoints. Small circles (o) represent the identified photospheric footpoints of the corresponding loops. The yellow dashed lines represent the background regions for those loops. The maroon contour indicates the umbra–penumbra boundary as obtained from the HMI continuum image.}
    \label{fig:aia12553}
\end{figure}

\section{Loop diameter along the coronal fan loop}
\label{Appendix:fwhm}

\begin{figure}
    \centering
    \includegraphics[width=0.99\linewidth]{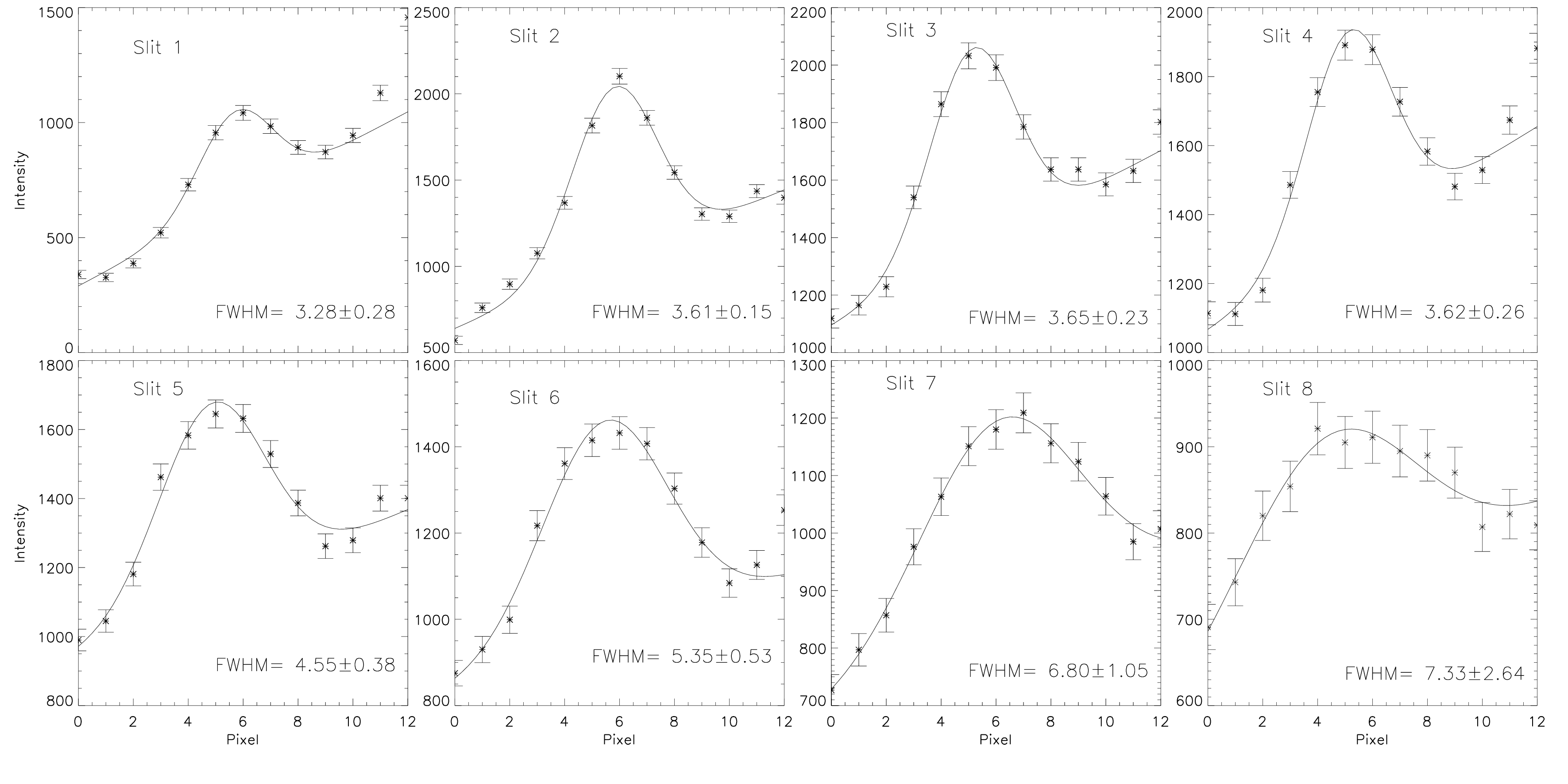}
    \caption{Intensity profiles along several slits across the coronal Loop 2 marked in Figure~\ref{fig:img} as observed in the AIA 171 \AA\ passband. All intensity profiles are fitted with the Gaussian function with a linear background. The obtained FWHM of the Gaussian function represents the diameter of the coronal loop along its length.}
    \label{fig:fwhm}
\end{figure}

\section{Details on the identification of loop locations and cross-sectional areas} \label{Appendix:area}

For the lower atmospheric heights, where the loops are not visible, recently \citet{2023MNRAS.525.4815R} devised a technique to trace back the footpoints and cross-section areas of coronal loops on the photosphere through the transition region and chromosphere. They studied fan loops emanating from sunspot umbra belonging to AR 12533 (7$^{\circ}$ S, $8^{\circ}$ N) observed simultaneously by IRIS and SDO on 2016 June 16. They utilized 4 hours of data starting from 07:19:11 UT. IRIS-SJIs recorded observations only in the 2796- and 1400-\AA\ passbands. However, due to poor signal, data from the 1400 \AA\ passband were not utilized. Therefore, they had very limited coverage over the whole solar atmosphere. In this study, along with AR 12553, we also utilize a new dataset belonging to the sunspot of AR 12470, which was observed in all four passbands of IRIS SJIs, along with SDO. Together, these passbands provide better coverage over the whole solar atmosphere (details in Section~\ref{sec:obs}). This makes the tracing of loops more robust in the lower atmosphere. Sunspots host waves and oscillations of a 3-min period in the umbral atmosphere and both 3-min and 5-min oscillations at the umbral photosphere \citep[for details see,][]{2015LRSP...12....6K}. \citet{2023MNRAS.525.4815R} utilized the presence of 3-min oscillations over the whole umbral atmosphere observed by the IRIS and SDO passbands to trace and identify the photospheric footpoints of coronal loops in the lower atmosphere through the transition region and chromosphere. They also measured the propagation speed of 3-min waves and found them to be less than the acoustic speed at all the atmospheric heights which confirmed that these are slow magnetoacoustic waves propagating along the traced loops. We also utilize the same technique to trace and identify the photospheric footpoints using a dataset with a larger number of passbands. This provides robust tracing of loops over the whole solar atmosphere through multiple coverages in the transition region and chromosphere. This also enabled us to deduce a better estimate of the loop cross-sectional areas at different heights in the umbral atmosphere.

\begin{figure}
    \centering
    \includegraphics[width=0.95\linewidth]{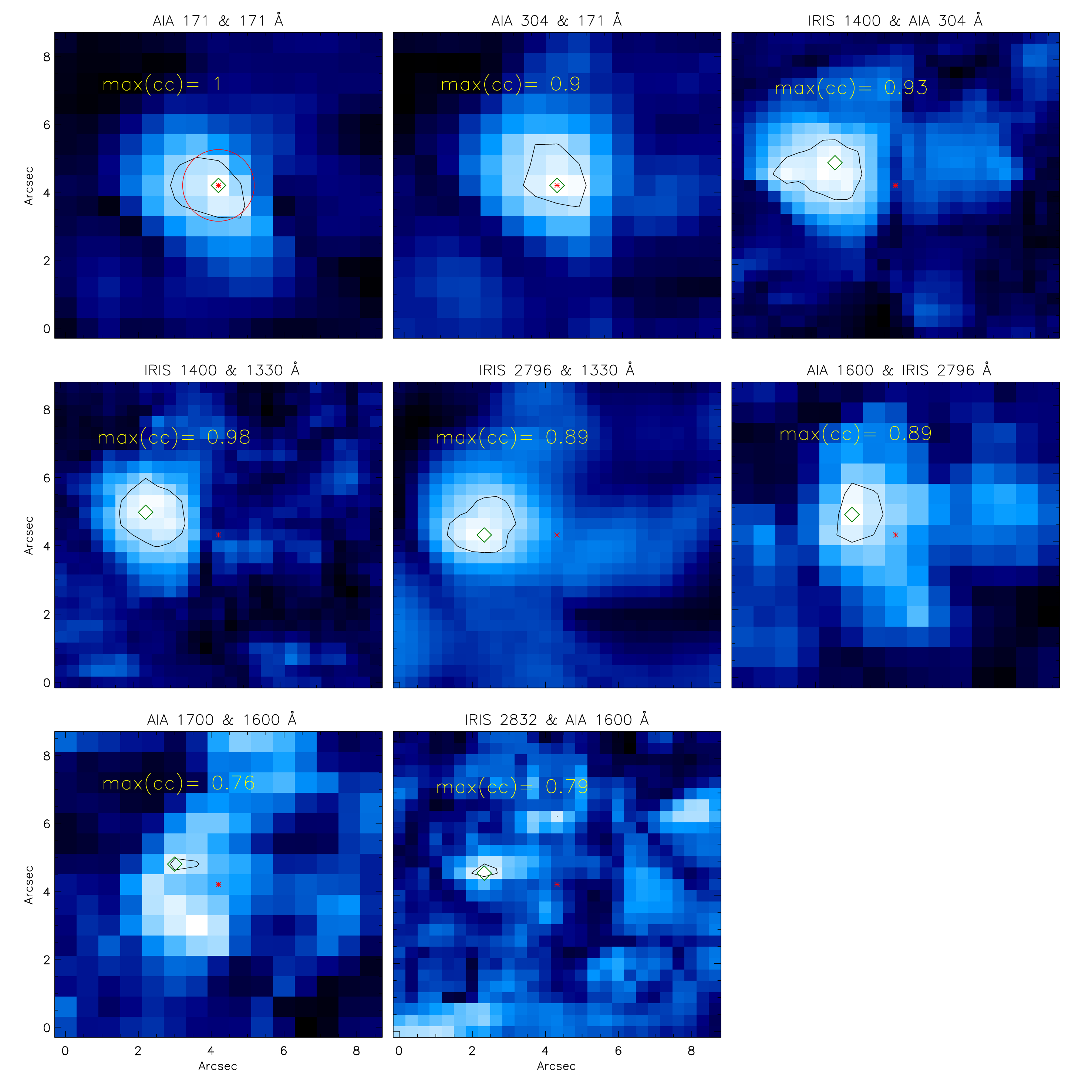}
    \caption{Correlation images obtained between various atmospheric heights as labeled. In each panel, the asterisk symbol (*) in the center refers to the coronal footpoint of Loop 2, and the red circle represents the cross-section of the loop obtained from the AIA 171 \AA\ image using the FWHM method. Overplotted black contours are obtained at $\approx94\pm2$\% of maximum correlation values. The green diamond ($\diamond$) symbols represent the locations of the maximum correlated pixel and the loop location at that height.}
    \label{fig:corr}
\end{figure}

As per the method described in \citet{2023MNRAS.525.4815R}, we identified several fan loops in the AIA 171 \AA\ image. We performed a correlation analysis to determine the loop locations and cross-sectional areas at different atmospheric heights in the lower atmosphere. We choose a box of size 9"$\times$9" shown by the yellow box in the AIA 171 \AA\ image in Figure~\ref{fig:img} by positioning the coronal footpoint of Loop 2 in the center of the box. We utilized 3-min (2-3.7 min period band) filtered light curves to perform the correlation analysis, which makes correlations depend only on 3-min oscillations. For correlating light curves between different passbands having different cadences, we first interpolated the light curves with longer cadences to light curves with smaller cadences using IDL routine $interpol$. We have first interpolated the IRIS 1400 \AA\ light curves to 12 s cadence to match the cadence of the 304 \AA\ passband and also interpolated the AIA 1600 \AA\ light curves to 12.75 s cadence to match the cadence of the IRIS 2796 and 2832 \AA\ passbands. We correlated the light curve at the coronal loop footpoint in AIA 171 \AA\ with light curves at each pixel in 9"$\times$9" box. The correlation image is then created by noting the maximum correlation coefficients at each pixel in the box. The pixel with the maximum correlation coefficient in the correlation image depicts the location of the loop at that atmospheric height. This location was further utilized to identify the loop location and cross-sectional area at the lower heights in the solar atmosphere by cross-correlating the light curves again with the 9"$\times$9" size box from the nearest temperature passband, as described above. In this way, we are able to trace the loop locations at different atmospheric heights till we locate their footpoint at the photosphere. The identified loop locations are marked with green diamonds ($\diamond$) in Figure~\ref{fig:corr} and are also marked by small circles in Figure~\ref{fig:img}.

In the first panel of Figure~\ref{fig:corr}, the overplotted red circle represents the cross-sectional area of the coronal loop footpoint in the AIA 171 \AA\ image as obtained from the FWHM method as shown in first panel (slit 1) of Figure~\ref{fig:fwhm}. Further, we chose a contour level on the correlation image by finding a value of the correlation coefficient at which the contour best fits the red circle. The obtained contour level for the loop footpoint in the corona is about 92\% of the maximum correlation coefficient value that fairly matches the loop cross-section obtained from the FWHM method. At various other heights in the lower atmosphere, a closed contour is achieved within $94\pm2$ \%  of the maximum correlation value level. These contours are overplotted in black color in all the correlation images in Figure~\ref{fig:corr}. These contours cover the correlation patches that are visible in the correlation images very well.

\end{document}